\documentclass[runningheads]{llncs}
\usepackage[T1]{fontenc}
\usepackage{graphicx}
\usepackage[hidelinks]{hyperref}
\usepackage{color}

\usepackage{listings}
\usepackage{tikz}
\usepackage{pgfplots}
\usepackage{pgfplotstable}
\usepackage{xcolor}
\pgfplotsset{compat=1.18} 
\usepackage{xfrac}
\usepackage{subcaption}
\usepackage{enumitem}
\usepackage{array, makecell}
\usepackage{amsmath}

\lstnewenvironment{cpptable}{\lstset{language=C++}}{}

\newcommand{\orc}[1]{\href{https://orcid.org/#1}{\includegraphics[height=10pt]{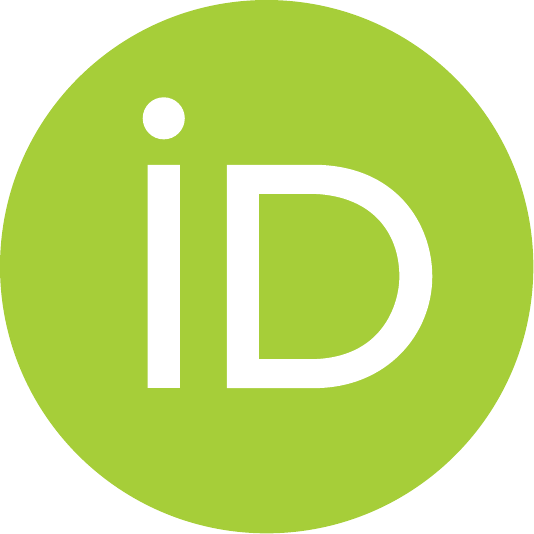}}}

\usepackage{wrapfig}

\begin{document}
\title{Closing a Source Complexity Gap between Chapel and HPX}
%
%
\author{Shreyas Atre\inst{1} \and
Chris Taylor\inst{2} \orc{0000-0001-7119-818X} \and
Patrick Diehl\inst{3}\orc{0000-0003-3922-8419} \and Hartmut Kaiser\inst{1}\orc{0000-0002-8712-2806}}
\authorrunning{S. Atre et al.}
%
\institute{Louisiana State University, Baton Rouge, LA, 70803 \and
Tactical Computing Labs, LLC, 
\and
Los Alamos National Laboratory, Los Alamos, NM, 87544\\
\email{\{satre1,hkaiser\}@lsu.edu,ctaylor@tactcomplabs.com,diehlpk@lanl.gov}}
\maketitle              
\begin{abstract}
A previous case study measured performance vs source-code complexity across multiple languages. The case study identified Chapel and HPX provide similar performance and code complexity. This paper is the result of initial steps toward closing the source-code complexity gap between Chapel and HPX by using a source-to-source compiler. The investigation assesses the single-machine performance of both Chapel and Chplx applications across Arm and x86. 


\end{abstract}

\section{Introduction}

HPX is an asynchronous many-task runtime (AMT) system implementing the ISO C\texttt{++} language's standard template library (STL) for data parallelism.HPX is built on the ParalleX execution model~\cite{parallex} placing emphasis onscalability, performance, and responsiveness of parallel applications using futures. HPX hosts a user-land thread library and an active global address space~\cite{hpx_agas}. 




Chapel~\cite{cascade} is a general purpose, productivity focused, procedural language designed with features supporting Partitioned Global Address Space (PGAS) programming~\cite{pgas}. Chapel gives users the ability to write applications under both a single-host shared memory system and a multi-host, distributed memory system. Chapel's language syntax shares similarities with C\texttt{++} and Python. Cray's Chapel compiler offers performance on par with C and C\texttt{++}.

The Cray Chapel compiler originally generated C programs. Chapel offers programming language features that were not organic to C. Support for the Chapel language specific features is provided by a custom C-based runtime system which this paper will refer to as Cray's Chapel runtime system. The current Cray Chapel compiler has shifted from source-to-source compilation to the LLVM compiler infrastructure. The Cray Chapel compiler continues to provide legacy support for C program generation.

The previous study~\cite{diehl2023benchmarking} indicated Chapel and HPX provided comparable performance. The study showed a gap between Chapel and HPX with respect to source code complexity. Source code complexity is measured using lines-of-code and the \textbf{Co}nstructive \textbf{Co}st \textbf{Mo}del (COCOMO)~\cite{Boehm1981} model. Closing the source-code complexity gap between Chapel and HPX requires implementation of a source to source compiler. The source-to-source compiler needs to translate Chapel software applications into ISO C\texttt{++} applications that exercise runtime system functionality provided by HPX.

 The ChplX compiler is the result of a 6 month software development effort. ChplX does not currently support the entire Chapel language. The ChplX compiler closes the code complexity gap between Chapel and HPX by achieving line-for-line translation of Chapel to C\texttt{++}. The ChplX compiler generates C\texttt{++} code that exercises functionality implemented in the ChplX library.

The ChplX library implements language functionality, using HPX, that was previously  exclusive to Chapel and the Chapel runtime system. The ChplX library implements facilities corresponding to Chapel data types like Chapel ranges, domains, arrays, but also Chapel language features like \textit{cobegin}, \textit{coforall}, and \textit{forall}, etc. The ChplX library provides ChplX compiler developers an ability to experiment with different code generation options. ISO C\texttt{++} developers have an opportunity to experiment with the ChplX library's implementation of the Chapel language's features outside of the Chapel programming language.

Our contributions in this work include the following:
\begin{itemize}
    \item A demonstration of Chapel's suitability as a frontend language for ISO C\texttt{++}.
    \item The first cross comparison of Chapel's performance across A64FX and x86.
    \item The ChplX\footnote{\url{https://github.com/ct-clmsn/chplx}} source-to-source proof of concept compiler and the ChplX C\texttt{++} library implemented using the AMT HPX~\cite{Kaiser2020}.
    \item A review of the Cray Chapel compiler's abstract syntax tree representation of conditional expressions and its impact on syntax driven management of variable scope.
    \item Prototyping a new C\texttt{++}-specific inlining feature for Chapel, `inlinecxx`.
    \item Evaluation of the strong scaling properties of 
    the Cray Chapel compiler compared to ChplX.
    \item An evaluation of the ChplX compiler's 6 month development effort to close the COCOMO gap while maintaining high performance.
\end{itemize}

This work demonstrates parity between the Chapel and HPX programming models. This work describes the ChplX compiler's implementation. The ChplX library and the functionality it provides is introduced. This work finishes with a performance evaluation of the 1D heat equation code, STREAM, GUPS, and code complexity using COCOMO scoring.

\section{The Chapel and HPX Programming Models}

Chapel provides a Single Program, Many Data (SPMD) distributed programming experience. Chapel offers a productivity oriented syntax emphasized in modern scripting languages. Chapel's syntax reflects its origins as an array programming language descended from Orca C and ZPL. Chapel extends its foundation as an array programming language by implementing a Partitioned Global Address Space (PGAS) communication model. PGAS has roots in UPC, Split-C, and OpenSHMEM~\cite{10.1145/2020373.2020375}.

HPX implements an ISO C\texttt{++} standard conforming data parallelism and asynchronous execution library. HPX introduces an Active Global Address Space (AGAS) built around its data parallelism foundation. Users can implement distributed object types that can expose methods and functions for remote invocation and execution. ISO C\texttt{++} futures become the foundation for managing the completion of remote method and function invocations. Users can chain multiple asynchronous invocations into a tree or dataflow graph.

HPX provides an implementation of C\texttt{++}20 coroutines along with the standard's new \textit{co\_await} functionality. HPX's coroutine support means C\texttt{++} is capable of performing Python and Chapel style yield and generator expressions. HPX's support for \textit{await} functionality makes the implementation of data parallel and recursive algorithms as straightforward as implementing a slightly modified version of the sequential version of the algorithm. Listing 5, in the Appendix, demonstrates the slight modifications required to parallelize a sequential implementation of the quick sort algorithm. HPX's coroutine support for C\texttt{++}20 coroutines creates the means to implement Chapel-style iterator functions. Chapel iterator functions are similar in form and function to Python's generator functions and expressions.

HPX and Chapel focus on shared-memory and distributed-memory parallelism and concurrency. Both provide abstractions for distributed data types and data structures. Both provide mechanisms for handling task dependencies and synchronization. Chapel and HPX share a common set of functionality and are different in a handful of ways.

HPX's AGAS implementation registers objects with the runtime using a string identifier. AGAS registered objects are capable of changing locality based on resource and compute constraints that may occur during an application's runtime. In contrast, Chapel fixes addresses within the partitioned global address space.

HPX provides first-class support for futures at the runtime system level. Chapel provides futures but, they are not an integral part of the underlying runtime system. Chapel's parallel model is fork-join in nature. HPX's model offers support for fork-join parallelism. HPX makes a concerted and deliberate effort to place futures and asynchrony at the forefront of its user story. The intent is to promote those tools over the classical fork-join model.

\section {The ChplX Toolchain and Environment}


The ChplX compiler part of the ChplX toolchain. The ChplX toolchain has three components. The first component is the ChplX compiler. The second component is the ChplX library. The third component is the Traveler visualization tools. The ChplX compiler creates C\texttt{++} application software. The compiler-generated application software uses the ChplX library. Users can opt to use the ChplX library without the ChplX compiler to craft C\texttt{++} applications. ChplX compiler and ChplX library users can both use and benefit from the Traveler visualization tools. The Traveler visualization tools exist to provide performance analytics for data collected from the HPX user-land thread library. Existing other performance tools are not compatible with HPX due the HPX user-land thread support.


\subsection{The ChplX Compiler}

The ChplX compiler performs the source-to-source translation of Chapel programs to C\texttt{++} programs. The ChplX compiler is implemented using C\texttt{++} and a frontend library provided by Cray's Chapel compiler. Cray's Chapel frontend library provides a lexer, parser, and abstract syntax tree framework implemented in C\texttt{++}.


The ChplX compiler adds directives in the generated C\texttt{++} program which point back to the original Chapel program. This feature helps debugging and visualization tools map the generated C\texttt{++} application software back to the original Chapel program. Users can quickly assess if a performance regressions or bottlenecks is the side effect of an implementation decision in the original Chapel program or an issue in the ChplX compiler generated C\texttt{++} program.

The ChplX compiler generates an appropriate CMake build file for the application. This feature offers a convenient mechanism to compile applications without having to bother with learning a build system or implementing build system scripts.

ChplX prototypes a Chapel function providing direct integration with C\texttt{++}. The function originally expedited the ChplX compiler's bootstrapping. The special function is called \textit{inlinecxx}.
The \textit{inlinecxx} function operates in a fashion similar to an \textit{asm} intrinsic found in C and C\texttt{++}. Users are able to write a string of C\texttt{++} code which is forwarded into the final generated program. The \textit{inlinecxx} function is implemented, in the ChplX compiler, using C\texttt{++}'s \textit{fmt} library. Listing 1 is an example of the feature.

\begin{lstlisting}[language=c++,caption=ChplX `inlinecxx` input and compiled C++ output,label=lst:rdtime,escapechar=|,float=tbp,showstringspaces=false]
// `inlinecxx` function signature
proc inlinecxx(string, n?...)

// Chapel code that uses `inlinecxx`
var i = 0;
inlinecxx("std::cout << i << std::endl");
inlinecxx("std::cout << {} << std::endl", i);

// ChplX generated C++ output from lines 5-6
int i = 0;
std::cout << i << std::endl;
\end{lstlisting}

Programs implemented using the ChplX \textit{inlinecxx} functionality will not be compatible with the Cray Chapel compiler. Users can correct this incompatibility by implementing the Chapel function in Listing 2 or by placing comments around their \textit{inline} expressions from their Chapel program. The function implemented in Listing 2 turns the \textit{inlinecxx} invocation into a "null" operation. The code provided to \textit{inlinecxx} passes through the Cray Chapel compiler with no effect on the output program. Depending on the complexity of the \textit{inlinecxx} workaround may not be sufficient for portability.

The ChplX compiler is oraganized into a frontend and backend. The frontend contains the parser and abstract syntax tree (AST) construction tooling provided by the Cray Chapel compiler. The backend accepts the AST created by the frontend and begins a multipass compilation process.

The first pass creates a \textit{program symbol table} using the visitor pattern\cite{gamma1994design}. The program symbol table is composed of smaller \textit{scope symbol tables} targeting each \textit{scope} (modules, functions, classes, loops, etc) structured using a heap.

The compiler populates scope symbol tables using Chapel's variable scoping rules. Each scope symbol table has parents and children. Scope symbol tables are organized as a heap and can be queried from the program symbol table using the line of code where the lexical scope is initialized. Program symbol table queries operate both up and down the tree.

Implementation of the symbol table pass is complicated by the lexical scoping structure of conditional expressions constructed by the Cray Chapel compiler's parser. The Chapel AST has a Block AST node representing a sequence of statements. Chained series of conditional expressions (\textit{if}, \textit{else-if}, \textit{else-if}) are nested inside each other in the AST as opposed to being constructed as a sequence of statements. Nested conditional expressions instantiate an excessive number of Block AST nodes increasing the complexity of the AST. Table 1 shows how a Chapel conditional expression is stored by the Cray Chapel compiler in the AST.

Variable declarations made within the scope of earlier conditional blocks can be naively propagated down into conditional block scopes. Syntactically this specific propagated variable declaration should be considered out of scope.

In situations where the final part of the conditional expression has no condition, an \textit{else} block, the nesting pattern is terminated. The \textit{else} block is added at the end of the conditional chain as a new, \textit{untyped block}. This particular structuring of chained conditional expressions and \textit{else} blocks complicates syntactic analysis for functions, classes, records, loops, conditional expressions, use of the AST for program symbol table construction, and enforcing variable scope. It is difficult to discern if the tail scope is an \textit{else block} or an extraneous conditional expression.

ChplX uses back pointers to manage these issues. There are some instances where back pointers causes unsatisfactory results. A better solution involves implementation of an initial compiler pass that transforms the Cray Chapel compiler's AST into a more manageable structure.

The initial transformation pass would create a new lexical type for conditional expressions. The new lexical type would store conditional expressions as variable length list of conditional clauses. The approach would more closely reflect the syntax reflected in the user's code and make managing scoping rules, symbol table creation, and symbol addition a straight forward process. This solution will be visited in future work.

\begin{table}[b]
\caption{Chapel language Conditional Expressions in the AST, `\{\}` represents a scope, `cond\(\)` represents the conditional\label{chplx_condexpr}}
\centering
\begin{tabular}{ l | l } 
Chapel Conditional & Chapel AST \\
\hline
\begin{cpptable}
if() {}
else if() {}
\end{cpptable}
&
\makecell[l]{
cond() \{ \\
\quad \{ \\
\quad\quad cond() \{ \\
\quad\quad \} \\
\quad \} \\
\}
}
\\
\end{tabular}
\end{table}


The ChplX symbol table is initialized with the following integral types \textit{nil, bool, string, int, real, complex, tuple, range, domain, and function}. The symbol table is additionally loaded with the template symbol (\texttt{?}), binary operators, ternary operator, and the \texttt{inlinecxx} function. Loops and conditionals are treated by the symbol table compiler pass as a special class of functions. Each instance of a loop or function is entered into the symbol table. Each loop or function is given a function name using the type of loop (\texttt{for}, \texttt{forall}, \texttt{coforall}, etc) or conditional expression (\texttt{if}, \texttt{else\_if}, \texttt{else}) along with the Chapel program's file name and the line number associated with the loop or conditional expression.

The second pass constructs data structures describing the output C\texttt{++} program using the program symbol table. Each AST node is visited and an output program syntax tree is constructed. The output program syntax tree uses \textit{std::variant} simplifying the traversal logic. Use of \textit{std::variant} reduces tree traversal to a \textit{std::visit} invocation. Output from the second pass creates a output program syntax tree. The output program syntax tree embeds symbol table information. This decision removes symbol table look ups from the code generation pass.

The final pass performs program code generation using the visitor pattern to print ISO C\texttt{++} to the output files. A CMake build script is generated. 

\begin{minipage}{0.45\textwidth}
\begin{lstlisting}[language=c++,caption=ChplX compiler input,label=lst:rdtime,escapechar=|]
var a : int = 1 + 1;
a = a + 1;
\end{lstlisting}
\end{minipage}
\hfill
\begin{minipage}{0.45\textwidth}
\begin{lstlisting}[language=c++,caption=ChplX compiler output,escapechar=|]
#line 8 "expr.chpl"
        auto a = 1 + 1;
#line 9 "expr.chpl"
        a = a + 1;
\end{lstlisting}
\end{minipage}


\subsection{ChplX Library}

Programs created by the ChplX compiler are composed by mapping Chapel's language syntax onto the ChplX library. The ChplX library is an abstraction layer above the HPX runtime system. ChplX implements Chapel's language functionality. The ChplX library is designed to simplify the source to source compilation process. The ChplX compiler's goal of achieving line-by-line translation is made possible by the implementation of the ChplX library.

The ChplX library currently provides support for: data parallelism, process level parallelism, range support, array support, tuples, domains, associative domains, atomics, locales, synchronization variables, and zippered iteration. The ChplX library can be extended further to provide generator function support, distributed classes and records, distributed containers, channels, and first class support for futures.

There are two benefits provided by the ChplX library. The first and more obvious benefit is that C\texttt{++} developers gain access to Chapel's language functionality. A secondary benefit is that the library provides C\texttt{++} developers an ability to further tune and optimize ChplX compiler generated Chapel programs.

The main design goal for the ChplX library is to reduce the complexity of transformations required to create ISO C\texttt{++} from Chapel. The library implements the main Chapel language semantics as C\texttt{++} template types such that there is almost a one-on-one relationship between a particular Chapel language feature and the corresponding ChplX library feature. Table~\ref{chplx_facilities} lists the most important facilities implemented by the ChplX library.

\begin{table}[tb]
\caption{Exemplar Chapel language features and corresponding ChplX library facilities\label{chplx_facilities}}
\centering
\begin{tabular}{ll|ll}
Chapel Language & ChplX Library & Chapel Language & ChplX Library  \\ 
\hline
Range           & \texttt{chplx::Range<>} & forall          & \texttt{chplx::forall}   \\
Tuple           & \texttt{chplx::Tuple<>} & coforall        & \texttt{chplx::coforall}    \\
Domain          & \texttt{chplx::Domain<>} & begin           & \texttt{chplx::begin}   \\
Array           & \texttt{chplx::Array<>} & cobegin         & \texttt{chplx::cobegin}  \\
Locale          & \texttt{chplx::Locale<>} & zip             & \texttt{chplx::zip}     \\
Dmap            & \texttt{chplx::dmap<>} &  Sync            & \texttt{chplx::Sync<>}  \\
Atomic          & \texttt{chplx::Atomic<>} & Single          & \texttt{chplx::Single<>}   \\
\end{tabular}
    \vspace{-5mm}
\end{table}

The APIs and semantics for the listed facilities enumerated in Table~\ref{chplx_facilities} are modeled after their corresponding Chapel language semantics. Using Chapel ranges as an example, the C\texttt{++} template type \texttt{chplx::Range<>} exposes all of the functionality defined for Chapel Ranges. ChplX Ranges support bounded and unbounded ranges, slicing, striding, arithmetic range operators, and the Range operators (\texttt{\#}, \texttt{by}, and \texttt{align}). All of the support is implemented as API functions operating on \texttt{chplx::Range<>} instances (\texttt{chplx::count()}, \texttt{chplx::by()}, and \texttt{chplx::align()}). \texttt{chplx::Range<>} additionally supports all of the member functions corresponding to the Chapel language specification. 


The parallel tasking and looping constructs are well integrated with the \texttt{chplx::Range<>}, \texttt{chplx::Domain<>}, and \texttt{chplx::Array<>} C\texttt{++} template types, such that iteration can be directly performed on instances of those.

Chapel iterators are directly mapped onto equivalent features exposed by the ChplX library. The implementation of iterators relies on native C\texttt{++} co-routines introduced in C\texttt{++}20, in particular on the C\texttt{++} language keywords \texttt{co\_return}, \texttt{co\_yield}, and \texttt{co\_await}. All iterable C\texttt{++} template types in the ChplX library expose specific iterators that enable seamless integration for the ChplX library looping and tasking constructs.

\section{Performance \& Complexity Evaluation}

Performance benchmark experiments were conducted on the following single node machines: Intel(R) Xeon(R) Platinum 8358 CPU clocked at 2.60GHz, with 64 cores, and 240 GiB DDR4 Memory and an A64FX machine with 32GB of HBM. All experiments were performed with Clang+LLVM 15, HPX v1.9.1 (APEX v2.6.3), Chapel 2.3, on an Ubuntu focal environment.

The following benchmarks are used to assess performance: STREAM, Heat equation, and GUPS. Each benchmark is compiled using different options. The first option exercises the Cray Chapel compiler's gcc and clang backends with the \textit{--fast} optimizations. The Cray Chapel compiler's llvm backend is not used in order to maintain a level of comparability between results. The second pair of options compile the Chplx compiler's generated code with gcc and clang using \textit{-O3} optimizations.



\subsection{Heat Equation Kernel}
%
The heat equation kernel evaluated in this study is one-dimension with length L for all times $t > 0$. Equation 1 defines the kernel. Equation 2
defines Euler's method which is used for time discretization. The $\boldsymbol{\alpha}$ term is a material's diffusivity.

While this problem is small and extremely simple, it has much in common with many high performance codes that simulate black holes, coastal waves, atmospheres, etc. Block-structured meshes that use finite differences are common and essential for modeling a wide variety of physical systems.

\begin{equation}
\frac{\delta u}{\delta t} = \alpha \frac{\delta ^ 2 u}{\delta \ u^2}, 0 <= x < L, t > 0
\end{equation}

\begin{equation}
u(t + \delta t, x_i) = u(t, x_i) + \delta t \cdot \alpha \frac{u(t, x_{i-1}) - 2 \cdot u (t, x_i) + u (t, x_{i+1}) }{2h}
\end{equation}

This specific kernel's performance has been studied across 10 programming languages and runtime systems in the previous study~\cite{diehl2023benchmarking}. The heat equation kernel was selected for this paper based based on the previous study's rationale. The problem is small, simple, and has commonality with high performance scientific codes in astrophysics~\cite{Diehl2023EvaluatingHA}, hydrodynamics~\cite{10.1007/s10915-019-00960-z}, and n-body simulations~\cite{Dekate_2012}.

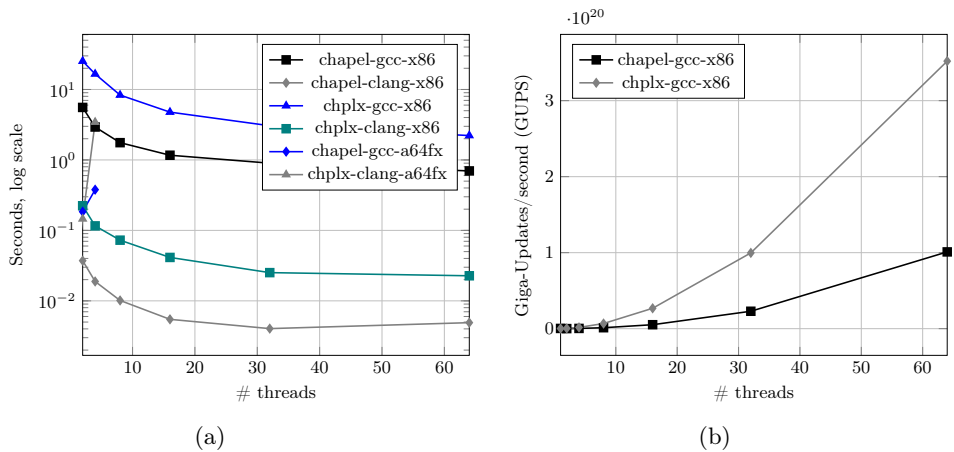
\begin{figure}[tbp]
\begin{subfigure}[t]{0.45\textwidth}
\begin{tikzpicture}[scale=0.75, transform shape]

\begin{axis}[grid,ymode=log,xlabel={\# threads},ylabel={Seconds, log scale},xmin=2,xmax=64, legend pos=north east]
    \addplot[black,thick,mark=square*] table [x expr=\thisrowno{0},y expr=\thisrowno{1}, col sep=comma] {2025_jan/heat-chapel-gcc.csv};
    
    \addplot[gray,thick,mark=diamond*] table [x expr=\thisrowno{0},y expr=\thisrowno{1}, col sep=comma] {2025_jan/heat-chapel-llvm.csv};
    
    \addplot[blue,thick,mark=triangle*] table [x expr=\thisrowno{0},y expr=\thisrowno{1}, col sep=comma] {2025_jan/heat-chplx-gcc.csv};

    \addplot[teal,thick,mark=square*] table [x expr=\thisrowno{0},y expr=\thisrowno{1}, col sep=comma] {2025_jan/heat-chplx-llvm.csv};

    \addplot[blue,thick,mark=diamond*] table [x expr=\thisrowno{0},y expr=\thisrowno{1}, col sep=comma] {2025_jan/arm-heat-chapel-gcc.csv};

    \addplot[gray,thick,mark=triangle*] table [x expr=\thisrowno{0},y expr=\thisrowno{1}, col sep=comma] {2025_jan/arm-heat-chplx-gcc.csv};

    \legend{chapel-gcc-x86,chapel-clang-x86,chplx-gcc-x86,chplx-clang-x86,chapel-gcc-a64fx,chplx-clang-a64fx};
\end{axis}
\end{tikzpicture}
\caption{}
\label{}
\end{subfigure}
\hfill
\begin{subfigure}[t]{0.45\textwidth}
\begin{tikzpicture}[scale=0.75, transform shape]

\begin{axis}[grid,xlabel={\# threads},ylabel={Giga-Updates/second (GUPS)},xmin=1,xmax=64,legend pos=north west]
    \addplot[black,thick,mark=square*] table [x expr=\thisrowno{0},y expr=\thisrowno{2}, col sep=comma] {2025_jan/gups-chapel-gcc.csv};
    \addplot[gray,thick,mark=diamond*] table [x expr=\thisrowno{0},y expr=\thisrowno{2}, col sep=comma] {2025_jan/gups-chplx-gcc.csv};
    \legend{chapel-gcc-x86,chplx-gcc-x86}; 
\end{axis}
\end{tikzpicture}
\caption{}
\label{fig:performance:gupsStrongScaling}
\end{subfigure}
\caption{Heat Equation: Strong Scaling, Average Time and (\subref{fig:performance:gupsStrongScaling}) GUPS \(n\_threads*bytes \div seconds \div 1e9\), Giga-Updates Per Second, 1GB Strong Scaling}
\label{fig:performance:heatEqn}
\end{figure}

Figure~\ref{fig:performance:heatEqn} is shows strong scaling performance for the 1D heat equation code using a log scale due to the magnitude of the values. The input data set size consists of 1 million double precision floating point numbers. 

\subsection{GUPS}

The Giga Updates Per Second (GUPS) benchmark~\cite{osti_860347} shown in Figure~\ref{fig:performance:gupsStrongScaling} measures how frequently a machine can update randomly generated memory locations. GUPS inundates the memory system with read and write requests. GUPS measures the machine's ability to perform updates while managing the parallelism implementations present in both the Cray Chapel runtime system and the ChplX/HPX runtime system. This GUPS benchmark is performed over approximately 1 gigabyte of memory. GUPS accesses memory locations that are within the space of half the total systems memory. Memory locations need be within a specific range and have little correlation. Updates are performed after a memory access. The GUPS implementation performs an `xor` on a 64-bit integer with a literal value. The result is stored back into memory. The GUPS benchmark metrics are computed by taking the number of threads, multiplied by the number of bytes processed, and dividing by the time.



\subsection{STREAM}

The STREAM\cite{McCalpin1995,McCalpin2007} benchmark is designed to measure the amount of sustained memory bandwidth and computational performance of vector kernels. STREAM reflects operations performed by simple computational kernels. The STREAM implementation is performed over an array of a million double precision floating point numbers. This data set size was selected to maintain consistency with the Heat Equation kernel. The STREAM benchmark performs three operations: addition, scale, and copy operation in parallel over varying data set sizes and thread counts. The implementation used for the STREAM performance study is a derivative of what is currently provided with the Cray Chapel compiler. Slight modifications have been made from the original Cray Chapel compiler's source due to current limitations in the ChplX compiler.



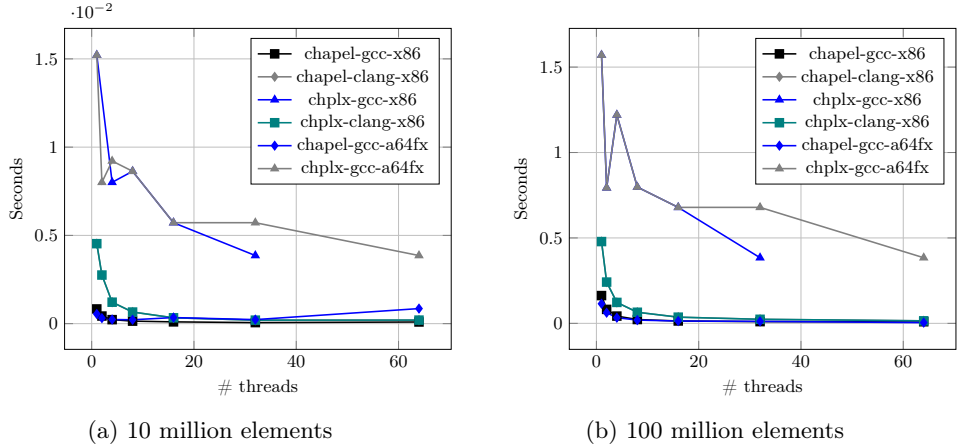
\begin{figure}[tb]
    \begin{subfigure}[t]{0.45\textwidth}
\begin{tikzpicture}[scale=0.75, transform shape]
\begin{axis}[grid,xlabel={\# threads},ylabel={Seconds},legend pos=north east]
    \addplot[black,thick,mark=square*] table [x expr=\thisrowno{0},y expr=\thisrowno{1}, col sep=comma] {2025_jan/stream-triad-chapel-gcc-1000000.csv};
    \addplot[gray,thick,mark=diamond*] table [x expr=\thisrowno{0},y expr=\thisrowno{1}, col sep=comma] {2025_jan/stream-triad-chapel-llvm-1000000.csv};    
    \addplot[blue,thick,mark=triangle*] table [x expr=\thisrowno{0},y expr=\thisrowno{1}, col sep=comma] {2025_jan/stream-triad-chplx-gcc-1000000.csv};
    \addplot[teal,thick,mark=square*] table [x expr=\thisrowno{0},y expr=\thisrowno{1}, col sep=comma] {2025_jan/stream-triad-chplx-llvm-1000000.csv};
    \addplot[blue,thick,mark=diamond*] table [x expr=\thisrowno{0},y expr=\thisrowno{1}, col sep=comma] {2025_jan/arm-triad-chapel-gcc-1000000.csv};  
    \addplot[gray,thick,mark=triangle*] table [x expr=\thisrowno{0},y expr=\thisrowno{1}, col sep=comma] {2025_jan/arm-triad-chplx-gcc-1000000.csv};
\legend{chapel-gcc-x86,chapel-clang-x86,chplx-gcc-x86,chplx-clang-x86,chapel-gcc-a64fx,chplx-gcc-a64fx} 
\end{axis}
\end{tikzpicture}
    \caption{10 million elements}
\label{fig:performance:stream_c}
    \end{subfigure}
    \hfill
    \begin{subfigure}[t]{0.45\textwidth}
    \begin{tikzpicture}[scale=0.75, transform shape]
    \begin{axis}[grid,xlabel={\# threads},ylabel={Seconds}, legend pos=north east]
    \addplot[black,thick,mark=square*] table [x expr=\thisrowno{0},y expr=\thisrowno{1}, col sep=comma] {2025_jan/stream-triad-chapel-gcc-100000000.csv};
    \addplot[gray,thick,mark=diamond*] table [x expr=\thisrowno{0},y expr=\thisrowno{1}, col sep=comma] {2025_jan/stream-triad-chapel-llvm-100000000.csv};    
    \addplot[blue,thick,mark=triangle*] table [x expr=\thisrowno{0},y expr=\thisrowno{1}, col sep=comma] {2025_jan/stream-triad-chplx-gcc-100000000.csv};
    \addplot[teal,thick,mark=square*] table [x expr=\thisrowno{0},y expr=\thisrowno{1}, col sep=comma] {2025_jan/stream-triad-chplx-llvm-100000000.csv};
    \addplot[blue,thick,mark=diamond*] table [x expr=\thisrowno{0},y expr=\thisrowno{1}, col sep=comma] {2025_jan/arm-triad-chapel-gcc-100000000.csv};
    \addplot[gray,thick,mark=triangle*] table [x expr=\thisrowno{0},y expr=\thisrowno{1}, col sep=comma] {2025_jan/arm-triad-chplx-gcc-100000000.csv};
\legend{chapel-gcc-x86,chapel-clang-x86,chplx-gcc-x86,chplx-clang-x86,chapel-gcc-a64fx,chplx-gcc-a64fx} 
\end{axis}
    \end{tikzpicture}

    \caption{100 million elements}
\label{fig:performance:stream_d}
    \end{subfigure}

    \caption{STREAM Benchmark}
    \label{fig:stream_bench}
\end{figure}

\subsection{Performance Summary}



ChplX demonstrates good performance and performance scaling based on the benchmarks presented in this study. Generally speaking, as data set sizes increase, Chplx performance increases. The ChplX C\texttt{++} library mirrors Chapel's features while maintaining high performance scalability. The ChplX compiler's emphasis on line-for-line source-to-source translation creates C\texttt{++} programs capable of maintaining competitive performance targets with respect to software compiled with the Cray Chapel compiler.



The heat equation kernel yielded Figure~\ref{fig:performance:heatEqn}. The kernels performance demonstrates the Cray Chapel compiler and the ChplX compiler have similar strong scaling profiles. The performance differences are believed to be a function of the runtime systems. Chplx on other benchmarks achieves improved performance scaling when the data set size exceeds a million elements. We believe that since the data set amount is fixed to 1 million elements similar performance improvements would be apparent as the data set size increases.

The STREAM benchmark in Figure ~\ref{fig:stream_bench} shows the Chplx/HPXruntime system offers performance that is comparable to the Cray Chapel compiler's runtime system. It's also clear that for large problem sizes Chplx can achieve performance targets consistent with those achieved by the Cray Chapel compiler.


The GUPS benchmark in Figure~\ref{fig:performance:gupsStrongScaling} shows a performance gap between ChplX and the Cray Chapel compiler. Chplx manages to exceed the Cray Chapel compiler's performance in this particular study across x86 and A64FX. We attribute this performance differential as a by product of Chplx's support libraries' use of HPX and the memory system.

\subsection{Source Complexity}

The ChplX compiler closes the complexity gap between Chapel and ISO C\texttt{++} HPX applications. The metrics used to define the code complexity gap are the COCOMO Estimated Scheduled Effort and Lines of Code. Closing the gap means there is an approximate line-for-line equivalency between Chapel programs and ChplX generated C\texttt{++} programs.

Figure~\ref{fig:loc} and ~\ref{fig:cocomo} show 3 frequencies between Chapel, C\texttt{++}, and Boilerplate over each benchmark. Chapel frequency measures the Chapel implementation of each benchmark under both metrics. ChplX generates several support files (header files, program driver files) along with the application specific program file. The application specific program file contains all the lines of code that directly correspond to the input Chapel program. C\texttt{++} frequency measures the C\texttt{++} application specific program file generated by ChplX. Boilerplate frequency captures measurements made with the application specific program file and the support files generated by ChplX.

The first quantitative measurement presents LoC measurements in Figure~\ref{fig:loc} using the cloc Linux tool. cloc has been directed to exclude comments and empty lines in the program files. The figure shows ChplX achieves a 5 line of code difference between all benchmarks.

The second quantitative measurement presents the COCOMO Estimated Schedule Effort. The previous heat equation study~\cite{diehl2023benchmarking} uses Estimated Schedule Effort (ESE). This paper exercises the same COCOMO metric for consistency. The ESE difference between each benchmark is ~0.7 ESE on average with a 0.6 median.

The differences between the Chapel Benchmark applications and the ChplX generated C\texttt{++} application software are marginal. A review of the software's generated source program shows the differences in the C\texttt{++} application software relate to the creation of namespaces and include directives. The program driver itself consists of a header file and program file and approximately 5 lines of software that initializes the HPX runtime system.

\pgfplotsset{compat=1.11,
    /pgfplots/ybar legend/.style={
    /pgfplots/legend image code/.code={%
       \draw[##1,/tikz/.cd,yshift=-0.25em]
        (0cm,0cm) rectangle (3pt,0.8em);},
   },
}

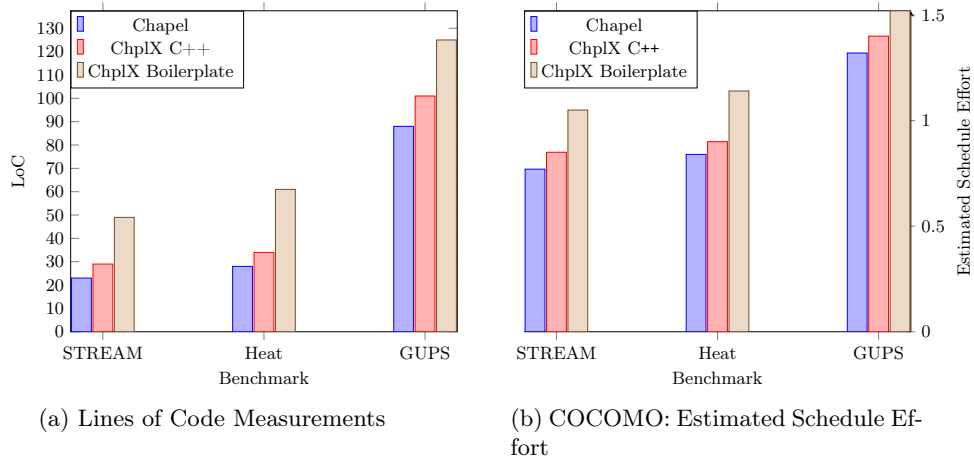
\begin{figure}[tb]
    \centering
    \begin{subfigure}[t]{0.45\textwidth}
        \pgfplotstableread[col sep=comma,]{loc_histogram.csv}\datatable
\begin{tikzpicture}[scale=0.75, transform shape]
\begin{axis}[
    ybar,
    xlabel={Benchmark},
    xtick=data,
    xticklabels from table={\datatable}{Benchmark},
    ybar=2*\pgflinewidth,
    ymin=0,
    scaled y ticks = false,
    ytick distance=10,
    ylabel={LoC},
    legend style={at={(0.0,1.0)}, anchor=north west}
    ]
    \addplot table[x expr=\coordindex, y={Chapel}] {\datatable};
    \addplot table[x expr=\coordindex, y={Cpp}] {\datatable};    
    \addplot table[x expr=\coordindex, y={Boilerplate}] {\datatable}; 
\legend{Chapel, ChplX C++, ChplX Boilerplate};
\end{axis}
\end{tikzpicture}
    \caption{Lines of Code Measurements}
    \label{fig:loc}
    \end{subfigure}
    \hfill
    \begin{subfigure}[t]{0.45\textwidth}
    \pgfplotstableread[col sep=comma,]{cocomo_data.csv}\datatable
\begin{tikzpicture}[scale=0.75, transform shape]
\begin{axis}[
    ybar,
    xlabel={Benchmark},
    xtick=data,
    xticklabels from table={\datatable}{Benchmark},
    ybar=2*\pgflinewidth,
    ymin=0,
    scaled y ticks = false,
    ylabel={Estimated Schedule Effort},
    legend style={at={(0.0,1.0)}, anchor=north west},
    axis y line=right
    ]
    \addplot table[x expr=\coordindex, y={Chapel}] {\datatable};
    \addplot table[x expr=\coordindex, y={Cpp}] {\datatable};    
    \addplot table[x expr=\coordindex, y={Boilerplate}] {\datatable}; 
\legend{Chapel, ChplX C\texttt{++}, ChplX Boilerplate};
\end{axis}
\end{tikzpicture}
    \caption{COCOMO: Estimated Schedule Effort}
    \label{fig:cocomo}
    \end{subfigure}
    \caption{Software estimates: (\subref{fig:loc}) Lines of code and (\subref{fig:cocomo}) estimated schedule effort using COCOMO.}
    \label{fig:enter-label}
        \vspace{-5mm}
\end{figure}

\section{Conclusions \& Future work}

The ChplX compiler demonstrates that source-to-source compilation to ISO C\texttt{++}20 combined with HPX, a standards conforming asynchronous many-task runtime system, can yield competitive high performance with the Cray Chapel compiler. The ChplX compiler is the end product of a 6 month collaboration yielding a solution that tightens the source code complexity gap between Chapel and ISO C\texttt{++}. The ChplX compiler prototyped and demonstrates the `inlinecxx` functionality works within the existing syntactic framework offered by Chapel. Development of the Chplx compiler identified a syntactic structural issue in the Chapel parsing infrastructure that complicates syntax driven compilation. The ChplX compiler shows Chapel's potential as a high performance frontend language for ISO C\texttt{++}. 

Future work will address ChplX compilation bugs that were discovered during implementation of the benchmarks. The performance gaps ChplX experiences will be studied and improved with experimentation with HPX's suite of executors. We intend to additionally investigate implementation of GPU support. Future studies can expand on this work by using additional benchmark codes to expand ChplX's language support. 

\section*{Acknowledgments}
\footnotesize
The authors would like to thank Stony Brook Research Computing and Cyberinfrastructure and the Institute for Advanced Computational Science at Stony Brook University for access to the innovative high-performance Ookami computing system, which was made possible by a \$5M National Science Foundation grant (\#1927880). This work was supported by the U.S. Department of Energy through the Los Alamos National Laboratory. Los Alamos National Laboratory is operated by Triad National Security, LLC, for the National Nuclear Security Administration of U.S. Department of Energy (Contract No. 89233218CNA000001). LA-UR-25-21252

%
%
%

%
%
%
%
\footnotesize
\bibliographystyle{plain}
\bibliography{chapel}

\end{document}